\documentclass[sigconf,screen]{acmart}
\copyrightyear{2019} 
\acmYear{2019} 
\acmConference[ESEC/FSE '19]{Proceedings of the 27th ACM Joint European Software Engineering Conference and Symposium on the Foundations of Software Engineering}{August 26--30, 2019}{Tallinn, Estonia}
\acmBooktitle{Proceedings of the 27th ACM Joint European Software Engineering Conference and Symposium on the Foundations of Software Engineering (ESEC/FSE '19), August 26--30, 2019, Tallinn, Estonia}
\acmPrice{15.00}
\acmDOI{10.1145/3338906.3340443}
\acmISBN{978-1-4503-5572-8/19/08}

 \titlenote{This work was funded by Vinnova, under the grant 2017-02963.}

\usepackage{graphicx}
\usepackage{amssymb}
\usepackage{enumitem}
\setlist[enumerate]{label*=\arabic*.}

\usepackage{rotating}
\usepackage{booktabs}
\usepackage{graphicx}

\usepackage{draftwatermark}
\SetWatermarkText{Pre-print}
\SetWatermarkScale{1}

\begin{document}

\title[Risks and Assets: A Qualitative Study]{Risks and Assets: A Qualitative Study of a Software Ecosystem in the Mining Industry}

\author{Thomas Olsson}
\affiliation{%
  \institution{RISE Research Institutes of Sweden AB}
  \streetaddress{Scheelev\"agen 17}
  \city{Lund}
  \country{Sweden}}
\email{thomas.olsson@ri.se}

\author{Ulrik Franke}
\affiliation{%
  \institution{RISE Research Institutes of Sweden AB}
  \streetaddress{Isafjordsgatan 22}
  \city{Kista}
  \country{Sweden}}
\email{ulrik.franke@ri.se}

\begin{abstract}
%% Text of abstract
Digitalization and servitization are impacting many domains, including the mining industry. As the equipment becomes connected and technical infrastructure evolves, business models and risk management need to adapt. 
%[Goal]
In this paper, we present a study on how changes in asset and risk distribution are evolving for the actors in a software ecosystem (SECO) and system-of-systems (SoS) around a mining operation. 
%[Method]
We have performed a survey to understand how Service Level Agreements (SLAs) -- a common mechanism for managing risk -- are used in other domains. Furthermore, we have performed a focus group study with companies. 
%[Results]
There is an overall trend in the mining industry to move the investment cost (CAPEX) from the mining operator to the vendors. Hence, the mining operator instead leases the equipment (as operational expense, OPEX) or even acquires a service. This change in business model impacts operation, as knowledge is moved from the mining operator to the suppliers. Furthermore, as the infrastructure becomes more complex, this implies that the mining operator is more and more reliant on the suppliers for the operation and maintenance. As this change is still in an early stage, there is no formalized risk management, e.g. through SLAs, in place. Rather, at present, the companies in the ecosystem rely more on trust and the incentives created by the promise of mutual future benefits of innovation activities.
%Conclusion
We believe there is a need to better understand how to manage risk in SECO as it is established and evolves. At the same time, in a SECO, the focus is on cooperation and innovation, the companies do not have incentives to address this unless there is an incident. Therefore, industry need, we believe, help in systematically understanding risk and defining quality aspects such as reliability and performance in the new business environment. 
\end{abstract}




%% keywords here, in the form: keyword \sep keyword

%% MSC codes here, in the form: \MSC code \sep code
%% or \MSC[2008] code \sep code (2000 is the default)

%%
%% Start line numbering here if you want
%%
%\linenumbers

%% main text

\maketitle

\section{Introduction}\label{Sec:Intro}
In modern software development and IT operations, cooperation and business partnerships are becoming more and more open -- through open source software, open innovation and open data. In the PIMM DMA project\footnote{www.sics.se/projects/pimm\label{footnote:PIMM}} -- a ``digital mining arena'' -- new business models and ways of cooperating are explored, including how to ``specify service level agreements\ldots in new value networks'' -- essentially handling risk. In the new service economy, with an increasing number of open collaborations, innovation and new business models become more important~\citep{chesbrough2006open} -- essentially asset distribution. However, it is also important for businesses to have predictability in costs and risk -- hence, there is still a need for agreements, formal or informal. In practice, there is usually a hybrid of open collaborations (sometimes referred to as software ecosystems (SECO)~\citep{axelsson2014characteristics,Messerschmitt2003}) and traditional ones regulated in contracts and agreements.

PIMM DMA (Pilot for Industrial Mobile communication in Mining, Digitalised Mining Arena) aims to foster innovative ideas and to increase the maturity level where industrialization can start and business relations can be initiated. Furthermore, PIMM DMA aims to build an understanding of how to specify services level agreements between involved suppliers and users, in new value networks. Innovation is focused to the following areas: 
\begin{itemize}
    \item Innovation in service operations for industrial mobile networks
    \item Innovation in cellular communication technology
    \item Innovation in industrial products and services enabled by cellular communication 
    \item Innovation in industrial automation focusing mining applications 
    \item Innovation in systems-of-systems
\end{itemize}

Examples of software-intensive services tested in the project include:
\begin{itemize}
    \item Connected Mining Operators
    \item Remote controlled wheel loader in production
    \item Connected Drill rigs
    \item Service Level Agreements
    \item AI and ML for realtime QoS prediction
    \item Process for building mobile networks in daily operations
\end{itemize}

The prototype implementations are tried out in an operational mine. 

In our previous work, we studied how technical infrastructure and business relationships are affected by System-of-Systems scenarios~\citep{borg2017}. The characteristics of how companies cooperate in the PIMM ecosystem influence risk management and innovation. For example, in traditional bilateral business relationships, risk is known and regulated in contracts. However, in the traditional buyer-supplier relationship, innovation might be weaker~\citep{borg2017}. On the other hand, if the companies collaborate in a SECO style with less regulated relationships, the dependency -- and thereby the risk -- can be more difficult to manage. A SECO, however, is one way to promote open innovation~\citep{chesbrough2006open} 

In this study, we focus on the distribution of risk and assets within a SECO. Companies manage risk, e.g., by agreeing on Service Level Agreements (SLAs), by insurances, or by moving asset risk to partners. Assets can either be acquired -- resulting in a capital expenditure -- or rented -- resulting in operational expenditure. Even though the latter means the company does not own the asset, the risk related to the asset is transferred to someone else. 

In a first step, we survey the prevalence of SLAs among -- and within -- companies. In a second step, we explore risk and asset distribution within the PIMM DMA ecosystem. More specifically, three research questions are investigated:

\begin{enumerate}
    \item[RQ1] What are the characteristics of risk management and SLAs in a software ecosystem in general and specifically in the PIMM DMA ecosystem?
    \item[RQ2] How are risks and assets distributed among the actors in the PIMM DMA ecosystem? 
    \item[RQ3] Which are the bridges and barriers when digitalizing a mining operation? 
\end{enumerate}

The rest of the paper is organized as follows: Section~\ref{Sec:BGRW} sets our work in context by discussing some related work. Section~\ref{Sec:Method} explains the research method, including a description of the respondents and PIMM DMA ecosystem. The main results are then presented in Section~\ref{Sec:Results} and discussed in Section~\ref{Sec:Discussion}. Finally, Section~\ref{Sec:Conclusion} concludes the paper.

\section{Background and Related Work}\label{Sec:BGRW}

Much of the current SLA research is technology-focused. For instance, many contributions aim to enable or improve automatic SLA negotiation between ``intelligent'' autonomous agents on interactive market places~\citep{silaghi2010framework,yaqub2014optimal}. Another strand of research addresses how service providers can maximize profits within the constraints imposed by the service levels~\citep{aib2007leveraging,casalicchio2013autonomic,goudarzi2012sla}. A good survey of SLA management focusing on these and other technical aspects is provided by \citet{Faniyi:2015:SRS:2856149.2843890}.

While these aspects are certainly relevant also in the kind of ecosystems that we address, our research has a different focus that also acknowledges \emph{informal negotiations} and \emph{human decision-making} as important aspects of SLAs. These aspects can be seen as setting the scene for the more technical ones -- before autonomous agents start to negotiate terms with each other, business decisions about selling, or procuring services in the first place must be taken. At the same time, it is known that the informal and human aspects pose problems, e.g., that IT departments fail to express availability service levels in ways that are understandable to the business side~\citep{gartner-bandwidth}, that value-maximizing SLA decisions are difficult even to professionals~\citep{franke2016experimental}, and that the information expressed in SLAs can lead to sub-optimal decisions~\citep{kieninger2013leveraging}.

Furthermore, recommendations on what to include in SLAs are not always consistent. For example, in principle it is known that it is prudent to express availability service levels both in terms of (i) total unplanned downtime allowed either as a percentage (e.g. 99.98 \%) or a number of hours (e.g.~10 h) , (ii) the maximum \emph{number} of unplanned outages allowed and (iii) the maximum \emph{duration} of any single unplanned outage allowed~\citep{franke2012optimal}. However, recommendations vary. Renowned consultancy Gartner sometimes emphasizes posing requirements on maximum outage duration~\citep{gartner-cost} but sometimes uses only a percentage~\citep{gartner-contactcenter} when recommending how to write SLAs. ITIL has an SLA template with both a percentage and a maximum number of unplanned outages, but without a maximum outage duration~\citep{itil-service-design}. As shown in our SLA survey (Section~\ref{Sec:pre-study}), similar variability is observed in practice.

SECOs changing how the software business is conducted~\citep{Messerschmitt2003}. In SECOs, the boundaries between supplier and customer are less clear, and companies that used to be competitors might choose to cooperate for mutual benefits~\citep{jansen2013software,axelsson2014characteristics}. SECOs is one way to promote open innovation~\citep{chesbrough2006open}. Typical examples are Google Android, Apache and Eclipse Foundation, which have established a novel approach to software engineering. The SECO approach refers to the collaboration of software activities across organizational boundaries, e.g. through (a) common software technology, appearing either as a platform, standard, or solution, (b) business, as a set of needs, possibly beyond profit and revenue, and (c) connecting relationships, as a community of actors~\citep{manikas2013software}. In a SECO, where relationships are not (solely) pecuniary, handling expectations such as on service levels changes compared to a more traditional B2B relationship. The cooperation are not necessarily pecuniary and powers of influence change. They can be both coercive e.g. how Google is controlling the API levels for approved Android phones or non-coercive such as expert or referent power~\citep{valencca2017theory}.

To summarize the related work, the predominant model of studying SLAs has been the technical point of view. Though it has been pointed out in the literature that SLAs should be described as a mixture of human-mediated functionality and computer-interpretable factors~\citep{Nguyen2017Aligning}, studies of SLAs from the non-technical perspective are scarce. Thus our work contributes to this important but under-investigated field, with a particular focus on human decision-making and business aspects of SLAs in an open innovation environment.

\section{Research Method}\label{Sec:Method}

\begin{figure*}[bt]
\includegraphics[width=0.75\textwidth]{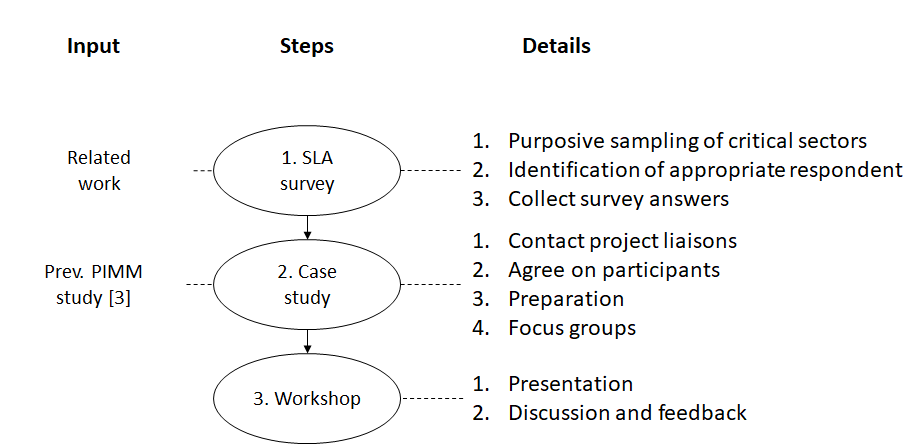} 
 \caption{Overview of our research process. }
 \label{fig:method}
\end{figure*}

We performed both a survey and a case study to address the research questions. In the first step, we surveyed enterprises in Sweden to understand the management of risk using SLAs, see Fig.~\ref{fig:method}. In the second step, we performed a case study in the PIMM DMA ecosystem to explore the risk and asset distribution among the companies. 

\subsection{SLA Survey \label{Sec:pre-study}}

In a first exploratory survey, we investigated how SLA practices differ depending on whether the SLAs govern IT services procured \emph{internally} (e.g. from an IT department) or \emph{externally} (e.g. from an IT consultancy). Such differences are interesting from the point of view of SECOs, as relations in ecosystems in a sense can be seen as an intermediate form, sharing some characteristics with internally procured IT services and some with externally procured ones~\citep{Messerschmitt2003}. 

These differences were explored with a questionnaire (see Appendix~\ref{app:Questionnaire}) asking the respondents first whether SLAs are at all used to govern the quality of IT services bought within the enterprise or from outside of the enterprise, respectively. If SLAs are used, the respondents were also asked whether they typically contain (i)~maximum unplanned downtime allowed, (ii)~maximum number of unplanned outages allowed, (iii)~maximum duration of an unplanned outage allowed, and (iv)~fines upon violation.

We are not aiming to characterize any particular population as a whole. Rather, we want to explore and understand of SLAs. Therefore, \emph{purposive sampling}~\citep{robson2011a} was employed to cover respondents representing sectors in society deemed particularly important by the Swedish Civil Contingencies Agency\footnote{www.msb.se}. The answers were collected anonymously. 

The survey answers are mainly analyzed by descriptive statistics. 

The survey was executed in February and March 2017. Respondents from four sectors participated: financial services ($N=2$) transport companies ($N=11$), food companies ($N=9$), and government organizations ($N=19$), i.e.~41 answers in total, depicted in Table~\ref{Tab:survey_participants}. (2 enterprises provided spuriously low revenue figures in the survey, so they are excluded from the analysis, and their revenues are listed as unknown in the summary in Table~\ref{Tab:survey_participants} and are left blank in Table~\ref{tab:pre-study-results}.)

\begin{table}[]
\caption{Participating enterprises in the survey, per domain and revenue size (Large $> 1000$ MSEK, Medium = 100 MSEK-1000 MSEK, Small $< 100$ MSEK).}
\label{Tab:survey_participants}
\begin{tabular}{l|llll|l}
Domain            & Large & Medium & Small & Unknown\footnote{Excluded} & All \\ \hline
Financial service & 1                                                                      & 1                                                                 &   
                                               &
                                                   & 2   \\
Food              & 3                                                                      & 5                                                                 & 1   
                                                &
                                                    & 9   \\
Government        & 9                                                                      & 8                                                                 & 1
                                                & 1
                                                    & 19  \\
Transportation    &                                                                        & 3                                                                 & 7
                                                & 1
                                                    & 11  \\ \hline
Total             & 13                                                                     & 17                                                                & 9                                                             &2
                                                    & 41 
\end{tabular}
\end{table}
\footnotetext{The organizations with unknown revenue size are not included in the analysis.}

\subsection{PIMM DMA Case study \label{Sec:interviews}}

The second part is an embedded case study~\citep{Runeson2012} of the PIMM DMA ecosystem. With the case study, we wanted to get a deeper insight into the SECO and the reasoning of the companies. The case study is executed as a focus group study. We carried out semi-structured focus groups~\citep{robson2011a} with the five companies from the PIMM DMA ecosystem. We performed focus groups with 1 to 4 respondents from each company. The focus groups were semi-structured in the sense that data was collected using a number of predefined but open questions~\citep{seaman2008qualitative}, see Appendix~\ref{app:InterviewGuide}.

All meetings were conducted as physical meetings with two researchers present -- one primarily taking notes and one primarily moderating the focus group session. 

All focus groups started with broad opening questions about current and future SLAs. Depending on the answers, we followed up with more questions from the list. The participants were encouraged to draw system descriptions on a whiteboard or similar to convey their answers and get a common understanding of the PIMM DMA ecosystem. The sessions ended with a wrap-up where we summarized the main findings and gave the opportunity to correct or add things.

After each focus group, of author wrote a summary, based on notes and memory, within one day. The other one then reviewed the summary and complemented the text. These notes, typically 1--2 regular pages of written text, were sent to the respondents. They were asked to review the notes and correct any misunderstandings.

Once the notes had been reviewed, they were coded. The 25 questions in the interview guide were used as input to create 33 codes. 
After initial coding by one researcher, the other researcher reviewed the result and codes were discussed until agreement was reached. 
Once all interview notes had been coded, selective coding was performed to identify the core codes and categories. Finally, core codes and categories were summarized to identify the main findings.

After analysis, representatives from all participating companies as well as other partners from the PIMM DMA project attended a seminar where the findings were presented. Everyone was given the opportunity to comment and correct on the findings as detailed in the presentation -- both those who were part of the focus groups as well as the others in the project.

The partners in the project are suppliers of products and services to a closed pit mine as well as the mining operator. The interviews were conducted during the winter of 2018/2019. The individual answers are confidential. Hence, in the text below, the companies are referred to by a randomized number. 

\section{Analysis and Results}\label{Sec:Results}

\subsection{Survey}
The results from the SLA survey indicate that SLAs are more prevalent when contracting external organizations (27 out of 39 respondents) compared to when cooperating across different parts of one company (21 out of 39), see Table~\ref{tab:SLAs}. Furthermore, larger organizations (as measured by revenue) are more prone to use SLAs than small organizations. Out of 10 enterprises with a revenue of less than 100 MSEK (small enterprises), just a single one uses SLAs internally and 2 use SLAs externally. As a contrast, out of the 13 enterprises with a revenue of more than 1000 MSEK (large), 8 use SLAs internally and 12 use SLAs externally. Of the medium-sized enterprises in our study, 13 of out 17 use SLAs externally and 11 out of 17 use SLAs internally.

\begin{table}[]
\caption{Prevalence of internal and external SLAs, per size and domain. (The complete data is available in Table~\ref{tab:pre-study-results}.)}
\label{tab:SLAs}
\begin{tabular}{@{}ll|ll|ll@{}}
       &                & \multicolumn{2}{l|}{External SLAs} & \multicolumn{2}{l}{Internal SLAs} \\
Size   & Domain         & Yes              & No              & Yes              & No             \\ \midrule
Large  & Financial      & 1                &                 &                 & 1              \\
       & Food           & 2                &    1             & 3                &                \\
       & Government     & 9               &                 & 5               & 4              \\ \midrule
Medium & Financial           & 1                &                & 1                &               \\
 & Food           & 3                & 2               & 3                & 2              \\
       & Government     & 8                &                 & 4                 & 4              \\
       & Transportation & 1                & 2               & 3                &               \\ \midrule
Small  & Food &                  & 1               &                  & 1              \\
  & Government & 1                  &                &                  & 1              \\
  & Transportation &    1              & 6               &   1               & 6              \\ \midrule
Total  &                & 27               & 12              & 20               & 19            
\end{tabular}
\end{table}

Finally, fines as a consequence for not fulfilling SLAs are quite common with external service providers (fines are used by 17 out of the 28 organizations using SLAs externally) but very rare with internal providers (fines are used only by 3 out of the 21 organizations using SLAs internally), see Table~\ref{tab:pre-study-results}.

\subsection{Focus Groups}

As seen in the previous section, external SLAs are common for large organizations but not necessarily smaller ones. Furthermore, internal SLAs are common though not to the same extent as external SLAs. We executed the focus groups to understand how this interplay works in a SECO mining context with different actors. The rest of the section summarizes the six major observations we made. 

\subsubsection{Transition to As-a-service Business Models}
The suppliers have different experience with selling machinery or services today. Supplier C focuses on selling systems and hardware, %D19
as does supplier A, %F19
though Supplier A is experimenting with selling operating-hours instead. %F20
Supplier B has a more diverse and customizable offer, where they sometimes sell hardware, sometimes use leasing agreements, and sometimes sell services such as operating-hours. %E19
Today, the mining operator mostly uses SLAs for ICT (Information and Communication Technologies) services, not operational technology used in the actual mining. %G19

Looking into the future, however, there is a clear overall trend that customers want to move from investments in equipment -- CAPEX -- to leasing or signing as-a-service agreements -- OPEX. This means that the cash flow changes as the suppliers take more of the upfront cost in exchange for receiving recurring payments for usage. On the other hand, the suppliers also take a bigger responsibility in the value-chain, where they are not only selling hardware but can also charge for maintenance and other services. % e19, d20, e20, f20, g20, D26

However, some suppliers are more skeptical as to whether it is possible to sign long enough contracts for the suppliers to be able to take the upfront cost; essentially, they are not sure that they want to assume that financial risk. %d26
Other suppliers manage such risks by clauses requiring the customer to buy a minimum number of operation-hours.%F22

On the customer side, a transition into the as-a-service paradigm also gives rise to new issues. For example, the mining operator wants to be able to audit and approve new software that is added even to machinery that it does not own. In case of new functionality, there is a willingness to pay for such updates, whereas corrective maintenance updates are expected to be included without additional cost until end-of-life.
%G26

\subsubsection{Economies of Scale}
Suppliers have equipment in several different mines, sometimes across the world. Having personnel to maintain the equipment and ensuring operation thus requires a local presence. Different suppliers have different approaches to this. 

Suppliers A and C work in line with what would be expected from the economies of scale point of view; focusing on standard products that are only somewhat customized to individual customers. Supplier C has remote monitoring and operation as part of their roadmap to be able to have a virtual presence in the mines without having to bear the full cost of personnel physically present. Supplier A is working extensively through local dealers and standard contracts and service levels, thus avoiding the need to have personnel of their own everywhere. %D20, F20

Supplier B, however, has chosen a different strategy, one that appears counter-intuitive from the economies of scale point of view: maintaining a strong local presence and extensively customizing the equipment to individual customers. Thus, development is more influenced by individual needs from the mines and less focused on off-the-shelf products. Still, supplier B has managed to maintain good profit margins  with this strategy. %E24, E26, E29

\subsubsection{Asymmetric Actors}
In a mine, there is a limited number of vehicles and devices. Even though a mining operator typically owns several mines, most of their suppliers operate on a global market. Hence, the suppliers have equipment in many mines owned by different companies. Furthermore, there are several different suppliers of equipment in a mine, even for similar or the same type of equipment. Some suppliers are very specialized, focusing on one part of the value chain whereas others cover larger parts of the value-chain for the kind of equipment they make.

Supplier B supplies equipment but is sometimes forced to take a systems integration responsibility even though they would prefer not to, e.g., for connectivity or integration with other systems. One reason is that they are physically present in the mines whereas providers of, e.g., connectivity, are not. % E5 in excel sheet.

Supplier C, however, provides more end-to-end integration of their equipment. Especially when it comes to data from the operation, they are skeptical to share and to use data from other organizations. Normally, the ``data integration'' is done in the control room by the personnel there. If the control room is no longer in the loop, not only will supplier C's position in the value chain be weaker, but the dynamic of decisions in the mine would change.  % D8 in excel

The mining operator, on the other hand, prefers to have a single point of contact if something is not working; they do not want to be the system integrator. However, they also realize that for, e.g., connectivity, the providers of both hardware and service are not present in the mine. Hence, they foresee that they will need to have employees or contractors locally to manage these services, as is already the case with some systems. %G24
This is especially true since the mine is a changing physical environment that is less static, compared to communication networks in other places. % G5, G10, g35

Suppliers A and B both try to keep up to date on ongoing technical standardization efforts, and to some extent also influence them. %E12, F12
This can be seen as a way to secure and ``future-proof'' their roles in emerging ecosystems.

\subsubsection{Partitioning of Risk}
The transition to as-a-service business models introduces more dependencies among the suppliers -- drilling-meters sold may depend on connectivity provided by another service provider. This translates into changes in the dynamics of maintenance and distribution of risk among the parties. Diagnostics will be different, requiring new competences and possibly new agreements to be able to live up to expected service levels. Consider, for example, a remote-controlled vehicle that requires connectivity. However, if the remote-controlled vehicle is not responding to commands, the root-cause can be found in several systems, not all of which are controlled by the vehicle supplier. %F35

For a supplier selling vehicle-hours as-a-service, a service level can only be guaranteed if it is clear who is responsible for connectivity, and on what conditions. %F21
Otherwise it is difficult to accept a contract where vehicle-hours are guaranteed on pain of monetary penalties to be paid. %G22
A possible alternative, raised by one supplier, is to find a way to make an SLA \emph{contingent} on, e.g., connectivity, i.e., contingent on the other actors delivering as they should. %F7

In general, as as-a-service business models become more commonplace, dependencies, roles, and partitioning of risks among the parties will need to be formalized in SLAs or similar contracts. %E20 %F20
The exact extent of this need, however, will depend on the architecture, and where integration take place. %D20
We expect that with the OPEX- and service-oriented paradigm, SLAs will become more important %F21
and the various ecosystem roles will become clearer. Depending on their risk appetites and business models, some actors might take on the role of an orchestrator or systems integrator, while others will want to avoid dependencies on others. %E29
In particular, the partitioning of risk across the actors will be a central concern. %F35 %G35

\subsubsection{Risks and Unintended Consequence of new Business Models}
Related to risk, supplier A makes the distinction between \emph{known} risks that can be mitigated, e.g., through prudent SLAs or business models, and the \emph{unknown} risks that can be introduced by new technology and new business models. For example, remote-controlled  vehicles introduce the risk of outages in connectivity, which can to some extent be mitigated through SLAs. However, they might also introduce more subtle behavior changes that are more difficult to plan for. For instance, as the driver of a remote-controlled vehicle is not in the vehicle, a crash or uncomfortable handling does not affect the driver in the same way. This might lead to different maintenance needs, as the equipment is worn differently. %F35

However, \emph{knowledge} of these risks may be unevenly distributed among the parties. For example, supplier B is able to analyze health data from some 2 000 machines globally, which none of their individual customers can do. %E13
Such knowledge may enable new roles, e.g., the ability to accept -- for a fee -- risks that are quantifiable and manageable to some, but not to all actors. %E13 %F13

\subsubsection{Sharing and Using Data}
All suppliers to the mine ecosystem agree that the operational data is owned by the mining operator. However, the mining operator does share this operational data with the suppliers. The security of any cloud solutions where such data is stored is paramount. %E35
Data is important to everyone to optimize their business. For example, data is used for preventive maintenance and to tune and optimize algorithms. There are no current business models based on data, however. The data integrity and confidentiality are handled on a contractual basis and not detailed in SLAs.  %D8, G4, E4, F24, F3, E3, G17
However, in a future with more exchange of data, additional contracts including SLA appendices may be necessary. %D35

Supplier C, however, focuses a lot on the quality of the data. For data to be used in operations they have to be able to trust it. Hence, they want to be able to control the entire process from collection, over transmission to processing of the data, and not allow it to flow through others' systems. %D29

\section{Discussion and Threats to Validity}\label{Sec:Discussion}

\subsection{RQ1 What are the characteristics of risk management and SLAs in a software ecosystem in general and specifically in the PIMM DMA ecosystem?}
Formal agreements on service levels are less common than we thought. In the survey, the large organizations use SLAs. Government organizations in particular, independent of size, have SLAs. We speculate that one possible explanation for government organizations using SLAs is that they, more than other organizations, make larger procurements including maintenance and operation.  

In the PIMM DMA case study, SLAs are uncommon. When we asked about this in the interviews, the answers were either that there is a kind of standard agreement or that there is a de facto service level which the organizations informally agree to. 

Ronald Coase famously asked why there are firms at all~\citep{coase1937nature}. Why is not everything bought on the market, using the price mechanism? His answer, in short, is that firms are established to avoid the transaction costs involved in buying things on the market rather than making them in-house. Among these are ``costs of negotiating and concluding a separate contract for each exchange transaction which takes place on a market''. We believe that more intertwined ecosystems possibly represent an intermediate form, not in-house but also not completely on the market.

Inspired by Coase, we thus hypothesize that in ecosystems that are relatively small and relatively durable, there is less need to regulate service levels formally in contracts. There is sufficient trust and alignment that the transaction costs of formally negotiating a contract can be saved. The PIMM DMA case is an example of this. However, in larger and more fluid ecosystems, there is a greater need to regulate service levels  formally in contracts. As relationships become more global and interaction and inter-dependencies among the organizations in the ecosystem increase in complexity, service levels will need to be managed  more formally than today. The comments from the PIMM DMA organizations interviewed about future developments point in this direction.

\subsection{RQ2 How are risks and assets distributed among the actors in the PIMM DMA ecosystem?}
As the domain is moving from a focus on CAPEX investments and owning the equipment to OPEX and instead leasing equipment or buying as-a-service, assets and risks move between organizations. Basically, the CAPEX investment is moved to the suppliers, which in turn are charging the mining operator for leasing or for a service, i.e., through OPEX.

In a sense, the capital risk, i.e., the risk of investing money upfront in development and construction that might not pay off in the end is thus moved from the mining operator to the suppliers. This implies, we believe, that the suppliers will take a greater responsibility for both the roadmap and innovation of new products and services as well as for the maintenance and continued operation. This, in turn, means that some knowledge is, ideally, moved away from the mining operator to the suppliers, or, worse, is lost on the way. In addition, we hypothesize that the position of the suppliers in the business relationship with the mining operator is strengthened. 

Technical risk, on the other hand, we believe is moved from the suppliers to the mining operator. As knowledge and operation is increasingly moved from the mining operator to the suppliers and the suppliers are dependent on each other and likely on sub-suppliers as well, overseeing and understanding the actual risk to the operation becomes more difficult. A failure in one service might halt the operation. If the relevant supplier cannot fix this failure quickly, the outage might be prolonged. Furthermore, maintenance is the responsibility of the suppliers, which means the mining operator are in the hand of the suppliers. As the operation becomes more and more dependent on services and on leased equipment, we believe that the technical risk to the operation becomes more complex to manage.

There is an ecosystem around the mining operation. The mining operator is the orchestrator and the different suppliers have different roles in the ecosystem. However, the different suppliers have their own ecosystems as well. Those consist of various components -- local or remote (in the ``cloud'') -- with additinal suppliers and subcontractors as well. These supplier ecosystems are orchestrated by them, and each of the different mines where the suppliers provide services can be seen as single actors in these supplier ecosystems. Hence, in essence, there are ecosystems of ecosystems: Mining operators orchestrate vertical integration of different services in their own mines, while their service providers orchestrate horizontal integration of their services, including product development and maintenance, across all the different mines where they operate. We believe this even further complicates the direct and indirect business relationships and how risk and assets are distributed. 

We hypothesize that as operations become more intertwined among ecosystems of ecosystems, there is an increasing need for risk management and risk awareness. Hence, we believe SLAs in these situations are relevant future work where software engineering and business research will need to meet. Furthermore, understanding of risk management on a higher level will also be needed. Sometimes, we believe, it is possible to estimate risks and consequences upfront, but not always. We believe that there is a need for further research in identifying the situations where the risk can be analyzed and when not. This implies that there is also a need to manage unknown risks, e.g. through cyberinsurances or other novel mechanisms in the ever more digitalized and software-intense mining industry. 

\subsection{RQ3 Which are the bridges and barriers when digitalizing a mining operations?}
We did not include detailed questions on data and data sharing in the interview guideline. However, the topic came up in all the focus groups anyway. Interestingly, there is a large consensus that the mining operator owns the data and that the data is sensitive from a business perspective. Still, the mining operator quite freely shares the data with the suppliers -- under non-disclosure agreements. Hence, the synergy is clear to all parties. We believe data is one bridge for further developing the digital ecosystem. It is also clear from the focus groups that the organizations all believe they are not yet using the data to its full potential. 

However, for data from other parties to be usable, data integrity and trustworthiness are paramount. Hence, there is a barrier today in that credible mechanisms for data validation are missing. This, we hypothesize, is both a technical barrier and a trust issue in the business relationships.

The transformation from a more traditional and CAPEX-oriented mining operation to a more digitalized and OPEX-oriented one is still early in its lifecycle. In fact, when we presented our findings to the project, they pointed out that the current technical infrastructure is still immature and evolving. Hence, the possible service level is difficult to estimate. Hence, in this environment, SLAs and formal agreements are difficult. Instead, the organizations accept the increased risk as this is still new and innovation focused. However, we believe at some point, when the operation has become sufficiently dependent on the service level on offer, SLAs will need to become more prevalent. At this point, we believe it will also be possible to pay more to get a better service level -- as the suppliers are competing and see that supplying a better service level is something they can charge for. However, we speculate that as the ecosystem matures and reaches the point where a high service level is no longer a competitive edge but a basic requirement, SLAs might again decline. This is the case, for example, for mobile telephony, where today we expect it to just work whereas a decade ago we might have been willing to pay extra for more bandwidth and shorter latency.

\subsection{Threats to Validity}
Here the validity of the conducted research is discussed based on commonly considered validity threats, e.g., as listed in \cite{robson2011a,Runeson2012}.

Construct validity concern the degree to which the constructs investigated are actually measured by the questions. For the survey part, construct validity is important to consider. The questions were iterated several times and tested internally with colleagues before sent to the respondents. Furthermore, we use standard terminology. Lastly, the research questions and overall scope was iterated several times among the authors as well as relevant stakeholders to assure our aim was right. In summary, we believe the threats to construct validity should be low. For the interview part, we performed a qualitative study with mostly open questions. Hence, construct validity is not a serious threat to the validity of the interview part. 

Reliability concerns the degree to which a respondent would give the same answers if they answered the same questions again under the same conditions. For the survey part, we used multiple choice questions and mostly asked questions which are objective and not personal opinion. Hence, threats to the reliability should be low. For the interview part, however, we used open questions and solicited qualitative answers. These type of answers are more personal opinion, depend more on the personal experience, what had happened the hours and days before the interviews (what is on the respondents' mind), etc. Hence, the threats to reliability for the interview part should be taken seriously. In essence, we do not believe the answers are, per se, wrong but might be incomplete. 

Internal validity is related to the causality of the study. However, we are not studying relationships among factors to draw casual conclusions. Rather, we are exploring the phenomena of SLAs. Hence, threat to internal validity are not a big concern, neither for the interview part nor the survey part. 

Regarding external validity (generalizability), we have not performed a statistically rigorous sampling approach (cf. Section~\ref{Sec:Method}). However, our aim is to understand the diversity and major trends, not make claims about the sampled population. Thus, as pointed out by~\citet{flyvbjerg2006a}, the threats to generalization should not be exaggerated. As we have covered different sizes, several domains, both technical and non-technical ones, and many locations in Sweden, we believe the findings to be, at least to some extent, relevant for software-intense companies in Sweden.  

\section{Conclusion and Future work}\label{Sec:Conclusion}
More and more of the equipment in a mine is connected. The result is an interconnected system-of-systems of physical equipment as well as cloud components -- from multiple suppliers. This impacts not only the technical infrastructure but also business models and risk management. 

The business model is moving from upfront CAPEX investments to servitization and OPEX cost instead. Hence, the mining operator is less and less owning equipment and relying more on the suppliers for operation and maintenance. The suppliers will take a larger responsibility for research and innovation, where previously the mining operator to a larger extent dictated what the suppliers should provide. 

As the mining operator is more reliant on suppliers for the operation, this impacts how risk is managed. Furthermore, as the technical infrastructure becomes more complex and the knowledge for how to maintain and handle problems is more specialized, the mining operator is increasingly dependent on the suppliers to ensure continued operation. SLAs is one way of managing risk through contractual agreements. Overall, SLAs are common -- at least for large companies -- in other domains. However, in the PIMM DMA context, this is not (yet) a practice. As the digilitzation of the mining operation is still in an early phase, it is difficult for the suppliers to commit to a certain level and for the mining operator to know which SLAs that are needed in the new system-of-systems. Hence, it seems, in order not to slow down innovation and introduction of new ways of working, the actors in the ecosystem rely on trust and mutual benefit. 

We see two main conclusions from our case study. On the one hand, SLAs are about managing risks and we believe that there is a need to better understand what the new technologies and business models implies for risk management. On the other hand, we speculate that in the companies studied in our case study, as long as there are no major incidents, there will be no incentives to improve the risk management in general and the SLA practices in particular. However, should there be an incident, then there will be a ``blame game'' and scramble to formalize risk distrbution in the ecosystem. If the PIMM DMA case is not unique, then the same implication is likely also valid for other ecosystems. 

The PIMM DMA project is one example of the ongoing digitilization and servitization which happens everywhere in society today. However, with risk and assets being distributed differently, we believe there is a need to understand the impact for the companies to ensure successful companies. Specifically, for complex ecosystems and systems-of-systems there is a need to better understand how to negotiate and agree to distribute risk. Furthermore, quality aspects such as reliability and performance will also be more difficult to assess. This, we believe, also requires more research to understand how such aspects can be evaluated upfront, during operation, and in the evolution of the system-of-systems. A critical aspect is to address the challenges and issues in industry and perform applied research and technology transfer.

%% The Appendices part is started with the command \appendix;
%% appendix sections are then done as normal sections
%% \appendix

%% \section{}
%% \label{}

%% References
%%
%% Following citation commands can be used in the body text:
%% Usage of \cite is as follows:
%%  ~\citep{key}          ==>>  [#]
%%   \cite[chap. 2]{key} ==>>  [#, chap. 2]
%%   \citet{key}         ==>>  Author [#]

% \pagebreak

\appendix

% Table generated by Excel2LaTeX from sheet 'Blad2'
\begin{table*}[h!]
  \centering
  \caption{SLA survey results on internal and external SLA usage by 41 respondents.}
  \resizebox*{!}{\dimexpr\textheight-3\baselineskip\relax}{%
    \begin{tabular}{rrr|rrrrr|rrrrr}
    \toprule
    \multicolumn{1}{l}{\begin{sideways}Category\end{sideways}} & \multicolumn{1}{l}{\begin{sideways}Employees:\end{sideways}} & \multicolumn{1}{l}{\begin{sideways}Revenue (MSEK):\end{sideways}} & \multicolumn{1}{l}{\begin{sideways}Uses internal SLAs\end{sideways}} & \multicolumn{1}{l}{\begin{sideways} \parbox{5cm}{\raggedright \ldots with maximum unplanned downtime allowed}\end{sideways}} & \multicolumn{1}{l}{\begin{sideways}\parbox{5cm}{\raggedright \ldots with maximum number of unplanned outages allowed}\end{sideways}} & \multicolumn{1}{l}{\begin{sideways}\parbox{5cm}{\raggedright \ldots with maximum duration of an unplanned outage allowed}\end{sideways}} & \multicolumn{1}{l}{\begin{sideways}\parbox{5cm}{\raggedright \ldots with fines upon violation}\end{sideways}} & \multicolumn{1}{l}{\begin{sideways}Uses external SLAs\end{sideways}} & \multicolumn{1}{l}{\begin{sideways}\parbox{5cm}{\raggedright \ldots with maximum unplanned downtime allowed}\end{sideways}} & \multicolumn{1}{l}{\begin{sideways}\parbox{5cm}{\raggedright \ldots with maximum number of unplanned outages allowed}\end{sideways}} & \multicolumn{1}{l}{\begin{sideways}\parbox{5cm}{\raggedright \ldots with maximum duration of an unplanned outage allowed}\end{sideways}} & \multicolumn{1}{l}{\begin{sideways}\parbox{5cm}{\raggedright \ldots with fines upon violation}\end{sideways}} \\
    \midrule
    Financial services & 250   & 1 200 &       &       &       &       &       & Yes   & Yes   & Yes   & Yes   & Yes \\
    Financial services & 400   & 900   & Yes   & Yes   &       &       &       & Yes   & Yes   &       &       &  \\
    Government & 7000  & 60 000 & Yes   & Yes   & Yes   & Yes   & Yes   & Yes   & Yes   & Yes   & Yes   & Yes \\
    Food  & 80    & 200   &       &       &       &       &       &       &       &       &       &  \\
    Food  & 55    & 80    &       &       &       &       &       &       &       &       &       &  \\
    Food  & 120   & 300   &       &       &       &       &       &       &       &       &       &  \\
    Food  & 150   & 278   & Yes   & Yes   &       &       &       & Yes   & Yes   &       &       &  \\
    Food  & 180   & 500   & Yes   &       &       & Yes   &       & Yes   &       &       & Yes   & Yes \\
    Food  & 220   & 2 000 & Yes   &       &       & Yes   &       &       &       &       &       &  \\
    Food  & 75    & 240   & Yes   & Yes   &       & Yes   &       & Yes   & Yes   &       & Yes   &  \\
    Government & 1200  & 3 400 &       &       &       &       &       & Yes   &       & Yes   &       & Yes \\
    Government & 1600  & 2 500 &       &       &       &       &       & Yes   & Yes   & Yes   & Yes   & Yes \\
    Government & 600   & 480   & Yes   & Yes   &       &       &       & Yes   & Yes   &       &       &  \\
    Government & 800   &      & Yes   & Yes   & Yes   & Yes   &       & Yes   & Yes   & Yes   &       & Yes \\
    Government & 55    & 60    &       &       &       &       &       & Yes   & Yes   &       &       & Yes \\
    Food  & 150   & 1 500 & Yes   & Yes   &       &       &       & Yes   & Yes   &       &       &  \\
    Government & 6000  & 4 000 & Yes   &       &       &       &       & Yes   & Yes   &       & Yes   & Yes \\
    Food  & 1000  & 3 000 & Yes   &       & Yes   &       &       & Yes   &       & Yes   &       &  \\
    Government & 1189  & 1 382 & Yes   & Yes   & Yes   & Yes   &       & Yes   & Yes   & Yes   & Yes   & Yes \\
    Government & 1100  & 6 000 & Yes   & Yes   & Yes   &       &       & Yes   & Yes   & Yes   &       &  \\
    Government & 2000  & 2 000 &       &       &       &       &       & Yes   & Yes   & Yes   & Yes   & Yes \\
    Government & 315   & 280   & Yes   &       &       & Yes   &       & Yes   & Yes   &       &       &  \\
    Government & 620   & 742   & Yes   &       &       & Yes   &       & Yes   &       & Yes   & Yes   & Yes \\
    Government & 550   & 525   & Yes   & Yes   &       & Yes   &       & Yes   & Yes   &       & Yes   &  \\
    Government & 5100  & 4 900 &       &       &       &       &       & Yes   & Yes   &       & Yes   & Yes \\
    Government & 1200  & 1 000 & Yes   & Yes   &       &       & Yes   & Yes   & Yes   &       &       & Yes \\
    Government & 135   & 137   &       &       &       &       &       & Yes   & Yes   & Yes   & Yes   & Yes \\
    Government & 850   & 906   &       &       &       &       &       & Yes   & Yes   &       &       & Yes \\
    Transportation & 60    & 75    &       &       &       &       &       &       &       &       &       &  \\
    Transportation & 80    & 80    &       &       &       &       &       &       &       &       &       &  \\
    Transportation & 60    & 250   & Yes   &       &       &       & Yes   &       &       &       &       &  \\
    Transportation & 31    &       &       &       &       &       &       &       &       &       &       &  \\
    Transportation & 60    & 48    &       &       &       &       &       &       &       &       &       &  \\
    Transportation & 44    & 80    &       &       &       &       &       &       &       &       &       &  \\
    Transportation & 37    & 35    &       &       &       &       &       &       &       &       &       &  \\
    Transportation & 57    & 77    & Yes   &       &       &       &       & Yes   &       &       &       &  \\
    Transportation & 200   & 300   & Yes   & Yes   &       &       &       &       &       &       &       &  \\
    Transportation & 42    & 32    &       &       &       &       &       &       &       &       &       &  \\
    Government & 1000  & 780   &       &       &       &       &       & Yes   & Yes   & Yes   &       & Yes \\
    Transportation & 100   & 200   & Yes   & Yes   &       &       &       & Yes   & Yes   &       & Yes   &  \\
    Government & 230   & 290   &       &       &       &       &       & Yes   & Yes   &       & Yes   & Yes \\
 \midrule
          &       & \textbf{Sum:} & \textbf{21} & \textbf{13} & \textbf{5} & \textbf{9} & \textbf{3} & \textbf{28} & \textbf{23} & \textbf{12} & \textbf{14} & \textbf{17} \\
    \bottomrule
    \end{tabular}%
    }
  \label{tab:pre-study-results}%
\end{table*}%

\section{Questions in the survey on SLAs}\label{app:Questionnaire}
\begin{enumerate}
    \item Type of enterprise (Private, government agency, county, municipality)
    \item Societal sector (Energy, financial, trade and industry, health and medical, information and communication, municipal technical services, food, public administration, protection and safety, social security, transport)
    \item Enterprise size (Numberof employees, revenue)
    \item Enterprise IT dependence (scale 1-10)
    \item Do you use SLAs to govern the quality of IT services bought within the enterprise, e.g. from an IT department? If yes, then please specify: 
    \begin{enumerate}
        \item Maximum unplanned downtime allowed
        \item Maximum number of unplanned outages allowed
        \item Maximum duration of an unplanned outage allowed
        \item Fines upon violation
    \end{enumerate}
    \item Do you use SLAs to govern the quality of IT services bought from outside of the enterprise, e.g. from an IT consultancy? If yes, then please specify: 
    \begin{enumerate}
        \item Maximum unplanned downtime allowed
        \item Maximum number of unplanned outages allowed
        \item Maximum duration of an unplanned outage allowed
        \item Fines upon violation
    \end{enumerate}
\end{enumerate}

\section{Interview guide}\label{app:InterviewGuide}
\begin{enumerate}
    \item Systems-of-Systems
    \begin{enumerate}
        \item Why is it created? 
        \item Why a SoS? 
        \item Whose system is it? 
        \item Who owns it? 
        \item Who are the stakeholders? 
        \item What should it do? 
        \item How much should it perform? 
        \item How should it be organized? 
        \item When does it change? 
    \end{enumerate}
    \item Software ecosystem
    \begin{enumerate}
    \item Governance (ad-hoc, democratic, autocratic)
    \item Role (orchestrator, niche, add-on)
\end{enumerate}
    \item Service Level Agreements
    \begin{enumerate}
        \item SLAs today
        \item SLAs missing
        \item Formality of SLAs
        \item Consequence if not fulfilling SLAs
        \item Frequency
        \item Rationale
        \item Origin
        \item Possibility to charge extra to get an SLA
        \item Bespoken or standard agreement
        \end{enumerate}
        \item Other
    \begin{enumerate}
        \item How is compliance communicated with partners? 
        \item Are there criteria on partners as such? 
        \item Will you have more or less SLAs in the future? 
        \item Which are the biggest challenges in terms of risk management? 
        \item What opportunities do you see with new business models and ways of working?
\end{enumerate}
    
\end{enumerate}

%% References with bibTeX database:

%\bibliographystyle{model1-num-names}
% \section*{References}
%\bibliographystyle{plainnat}
\bibliographystyle{ACM-Reference-Format}
\bibliography{references.bib}

%% Authors are advised to submit their bibtex database files. They are
%% requested to list a bibtex style file in the manuscript if they do
%% not want to use model1-num-names.bst.

%% References without bibTeX database:

% \begin{thebibliography}{00}

%% \bibitem must have the following form:
%%   \bibitem{key}\ldots 
%%

% \bibitem{}

% \end{thebibliography}

\end{document}